# Dimensional Confluence Algebra
# of Information Space Modulo Quotient Abstraction Relations
# in Automated Problem Solving Paradigm

*Author:* **Seppo Ilari Tirri**

## ABSTRACT

Confluence in abstract parallel category systems is established for net class-rewriting in iterative closed multilevel quotient graph structures with uncountable node arities by multi-dimensional transducer operations in topological metrics defined by alphabetically abstracting net block homomorphism. We obtain minimum prerequisites for the comprehensive connector pairs in a multitude dimensional rewriting closure generating confluence in Participatory algebra for different horizontal and vertical level projections modulo abstraction relations constituting formal semantics for confluence in information space. Participatory algebra with formal automata syntax in its entirety representing automated problem solving paradigm generates rich variety of multitude confluence harmonizers under each fundamental abstraction relation set, horizontal structure mapping and vertical process iteration cardinality.

## INTRODUCTION

This work is a general study of multitude confluence in multi-level generated information operator space and comprises abstraction structuring notation confluence of sets of participatory algebra operations regarded as formal semantics for information space. The elements in Participatory algebra stand for the most general representation of information units and operations in Participatory algebra as class rewriting represent alternations of knowledge. After representing necessary preliminary results and definitions from my previous works in Arxiv.org within automated problem solving paradigm the work is set to focus on constructing minimum prerequisites for connector pairs to harmonizing the initial arbitrarily chosen information rewriting operations later to be expanded confluence to cover also Cartesian cases of operations to be harmonized. Generation in Participatory algebra via its abstraction relation class rewriting operations representing automated problem solving paradigm generates rich variety of multitude confluence harmonizers liable to each fundamental abstraction relation set, horizontal structure mapping and ordinal ordered vertical process iteration.

.

## 1. §          Preliminaries

First we recall the central preliminaries of net representation and renetting systems as well as basic results of closures with abstraction relation in Tirri SI (2013 May) and Tirri SI (2013 Aug).

**Definition 1.1.** APPLICATION TYPES of RNS. For given RNS $\mathcal{R}$, jungle S is $\mathcal{R}$-*rewritten* to jungle T (*rewrite result*), denoted S $\rightarrow_{\mathcal{R}}$ T (called $\mathcal{R}$-*application*), and is *reduced under* $\mathcal{R}$ or by rule $\varphi$ in $\mathcal{R}$, and is said to be a *rewrite object* for $\mathcal{R}$ or $\varphi$ respectively, denoting T = S$\varphi$ (the postfix notation is prerequisite), if the following "rewrite" is fulfilled:

$$T = \bigcup (S(\mathbb{p} \leftarrow (\text{right}(r))g) : \text{left}(r) \text{ matches s in } \mathbb{p} \text{ by some net substitution mapping } f_{\mathbb{s}\mathbb{p}}, r \in \varphi, g \in G_{\mathbb{s}\mathbb{p}}, \mathbb{p} \in p(s), s \in S, \mathcal{C}(\mathcal{R})),$$

where $G_{\mathbb{s}\mathbb{p}}$´s are sets of net substitution relations. Mapping $f_{\mathbb{s}\mathbb{p}}$ is called *left side substitution relation* and each g in $G_{\mathbb{s}\mathbb{p}}$ is *right side substitution relation*, c.f. under conditional demands "extra variables on right-hand sides". We say that RNS is S-*instance sensitive* (S-INRNS), if for a rule $\varphi \in$ RNS and for each $s \in S$, $\mathtt{p} \in p(s)$, $G_{s\varphi} \neq f_{s\varphi}$.

Definition 1.2.  UNIVERSALLY PARTITIONING RNS. For each jungle (here c) we define a *universally partitioning RNS* (UPRNS) $\mathcal{W}$ of that jungle as a RNS fulfilling conditions (i)-(iii):

(i)    $\mathcal{W}$ is thoroughly totally environmentally saving and outward rank number saving,

(ii)   $\mathcal{C}(\mathcal{W}) \supseteq \{L(c) \cap L(c\mathcal{W}\hat{\ }) = \varnothing\}$,

(iii)  $L(\text{apex}(\text{right}(r))) \setminus \Xi$ is a singleton and its element is outside $L(c)$, whenever $r \in \varphi$, $\varphi \in \mathcal{W}$, and $\{(\text{left}(r), \text{right}(r)): r \in \varphi, \varphi \in \mathcal{W}\}$ is an injection.

Definition 1.3. N-th GRADE NET NUO-PRESENTATION, a linkage presentation.

For net $t = s(\mu_i; \lambda_j \mid i \in \mathscr{I}_s^{UN}, j \in \mathscr{J}_s^{UN}, C)$ set $\{s, \mu_{iL}, \lambda_{jL}: i \in \mathscr{I}_s^{UN}, j \in \mathscr{J}_s^{UN}\}$ is entitled *block* of t. Let then $T = \{s, \mu_i, \lambda_j: i \in \mathscr{I}, j \in \mathscr{J}\}$, where $\mathscr{I} \subseteq \mathscr{I}_s$, $\mathscr{J} \subseteq \mathscr{J}_s$ and the indexed nets in T are supposed to occupy indicated arity letters of t in s, we say that $s(\mu_i; \lambda_j \mid i \in \mathscr{I}, j \in \mathscr{J}, C)$, is a *net NUO-representation* of t and we denote $t = s(\mu_i; \lambda_j \mid i \in \mathscr{I}, j \in \mathscr{J}, C)$, and set$\{s, \mu_{iL}, \lambda_{jL}: i \in \mathscr{I}, j \in \mathscr{J}\}$ is entitled its *block*. NUO(t) is asserted on the establishment for the set of all NUO-representations of t. Notation block(t) stands for the family of all block-collections in NUO(t). Notice that each $\mu_{iL}$ and $\lambda_{jL}$ may be nets in enc(t) not necessarily totally isolated from s, although $\mathscr{I} \cap \mathscr{I}_s^{UN}$ and $\mathscr{J} \cap \mathscr{J}_s^{UN}$ may be nonempty, in other words $\mu_{iL}$ or $\lambda_{jL}$ may overlap net s thus comprising the key feature of NUO-representation. Observe also that $\{s, \mu_i, \lambda_j: i \in \mathscr{I}, j \in \mathscr{J}\}$ is a cover of t, and conversely for any cover of t there is such a NUO-representation that each element in the said cover stands for a net in the block of a net in NUO(t). For NUO(t)-representations classes of equal nets [s], $s \in$ NUO(t), are defined analogous with t-class definition. Furthermore we obtain block([t]) = $\cup$(block(p) : $p \in [t]$), where [t] is the set of t-class representatives. If net p is in NUO(enc([q])), then q = [q(p | )] = [p(q | )]. Because NUO-representation is covering the net definition, in the following our presumption (if not stated other) is simply to use NUO-representation for nets and assume indexes in nets be subsets of indexes in block elements.  For each net t we denote reverse NUO, NUO $^{-1}$ mapping, by asserting:  t = NUO$^{-1}$(NUO(t)). As usual for sets we define block(S) = {block(s) : $s \in$ S}.

We generalize NUO-presentation by induction to deal with n-th grade to serve as a more transparent view of "multidimensional" character of nets as follows:

First we denote the first grade NUO as $NUO^1$. Let $n \in IN$ and $s_n(\mu_{n,i}; \lambda_{n,j} \mid i \in \mathscr{I}_n, j \in \mathscr{J}_n, C_n)$ be a n-th grade NUO of v i.e. $NUO^n$-representation of v. Then

$s_n(\mu_{n,i}(\mu_{\mu,n+1,i}; \lambda_{\mu,n+1,j} \mid i \in \mathscr{I}_{\mu,n+1}, j \in \mathscr{J}_{\mu,n+1}, C_{\mu,n+1}) ; \lambda_{n,j}(\mu_{\lambda,n+1,i}; \lambda_{\lambda,n+1,j} \mid i \in \mathscr{I}_{\lambda,n+1}, j \in \mathscr{J}_{\lambda,n+1}, C_{\lambda,n+1}) \mid i \in \mathscr{I}_n, j \in \mathscr{J}_n, C_n)$

is a $NUO^{n+1}$-representation of v. In accordance we define reversed $NUO^{-n}$, $n \in IN$. For each $n \in IN$ we define $block^n(t) = \cup(block(enc(NUO^n(t))))$. $NUO^m NUO^n(t)$ is $NUO^m$ context set of $NUO^n(t)$, when $m \leq n$.

## Definition 1.4. N-th GRADE NET BLOCK HOMOMORPHISM.

Let $t = s(\mu_i; \lambda_j \mid i \in \mathscr{I}, j \in \mathscr{J}, C)$ , where $\mathscr{I} \subseteq \mathscr{I}_s$, $\mathscr{J} \subseteq \mathscr{J}_s$, be a net in $F_\Sigma(X, \Xi_\Sigma)$. We define *net D-block homomorphism relation* (D-NBH) h as net homomorphism, but ranked letter rewriting relation $h_\Sigma$ is replaced by *net D-block rewriting relation* $h_D : (D, I_D) \mapsto h_\Sigma(\Sigma)$ , $D \subseteq F_\Sigma(X, \Xi_\Sigma)$, $I_D$ is index set $\{ k : s_k \in D \} \cup \{ \varnothing \}$ with mapping $I_D^\wedge : s_k \mapsto k$  and

$$h(t) = h_D(s, I_D^\wedge(s))( h(\mu_i) ; h(\lambda_j) \mid i \in \mathscr{E}_{inh_D(s, I_D^\wedge(s))}{}^{UN}, j \in \mathscr{E}_{outh_D(s, I_D^\wedge(s))}{}^{UN}).$$

Jungle D is indicated via corresponding NBH h by notation block(h).

Relation h is *alphabetically unexpanding* (AlpUnexNBH), if $h(X \cup \Xi_\Sigma) \subseteq Y \cup \Xi_\Omega$, and for each $d \in D$ $\mid rank(h_D(d, I_D^\wedge(s))) \mid$ is not greater than $\mid rank(d) \mid$. Furthermore we say that h is entitled *right hand side alphabetical* (AlpNBH), if $h(X \cup \Xi_\Sigma) \subseteq Y \cup \Xi_\Omega$, and for each $\sigma \in \Sigma$  $h_\Sigma(\sigma) = \omega(\varepsilon_i ; \varepsilon_j \mid i \in \mathscr{E}_{in} , j \in \mathscr{E}_{out})$, where $\omega \in \Omega$ and $\{ \varepsilon_i ; \varepsilon_j \mid i \in \mathscr{E}_{in}, j \in \mathscr{E}_{out} \}$ is an arity alphabet. We say that it is *abstracting* (ANBH), if it does not delete the contexts (e.g. s in our example) of (overlapping) subnets (e.g. $\in \{ \mu_i, \lambda_j : i \in \mathscr{I}, j \in \mathscr{J} \}$) and preserves at least one linkage between the preimage contexts and each of their (overlapping) subnets. Right hand side alphabetical and environment saving net block homomorphism is called *alphabetically abstracting*, abbreviated AlpANBH. A special case of net block homomorphisms is *n-th grade* where is defined $D = block^n(F_\Sigma(X, \Xi_\Sigma))$, where domains of homomorphisms comprise $NUO^n$-representations.

We recall definitions of micro and macro of rules liable to intervening *renetting systems* (RNS) types and centers of corresponding abstraction relation and notice that the operational efficiency of the set of all *net block homomorphism* NBH´s and on the other hand of all RNS´s equates. Parallel *transducers* (TD´s) derived from those micro and macro form classes of *parallel TD equivalence relation* $\infty^c_\theta$ for θ-types of ITG (={PRNS,GPRNS,CLRNS,UPRNS}) in the set of all TD´s. Two jungles are in *NBH-abstraction relation* with each other, if they are NBH-images of NUO-presentations of the same jungle.

Let θ be *alphabetically abstracting net block homomorphism* AlpANBH-abstraction relation. Let (A,B) be an AlpANBH-abstract pair of nets, $\mathcal{W}_1$ and $\mathcal{W}_2$ being of AlpANBH-type intervening NBH´s in concern, A $\mathcal{W}_1$-image of net s and B $\mathcal{W}_2$-image of net t while s and t being *net NUO-representations* of the same net, say K, with block of s $D_s \subseteq$ block($\mathcal{W}_1$) and block of t $D_t \subseteq$ block($\mathcal{W}_2$). Without a loss of generality we make an assertion:

$s = p(\mu_i; \lambda_j \mid i \in \mathcal{S}_p, j \in \mathcal{G}_p, C)$ and

$t = q(\mu_i; \lambda_j \mid i \in \mathcal{S}_q, j \in \mathcal{G}_q, (\exists\, k \in \mathcal{S}_q^{oc} \cup \mathcal{G}_q^{oc})\, p \in \{\mu_k, \lambda_k\}, C)$.

## 2. §        Abstraction Parallel Transformations

Let θ be *alphabetically abstracting net block homomorphism* AlpANBH-abstraction relation. Two jungles are in *NBH-abstraction relation* with each other, if they are NBH-images of NUO-presentations of the same jungle. Let (A,B) be an AlpANBH-abstract pair of nets, $\mathcal{W}_1$ and $\mathcal{W}_2$ being of AlpANBH-type intervening NBH´s in concern, A $\mathcal{W}_1$-image of net s and B $\mathcal{W}_2$-image of net t while s and t being *net NUO-representations* of the same net, say K, with block of s $D_s \subseteq$ block($\mathcal{W}_1$) and block of t $D_t \subseteq$ block($\mathcal{W}_2$). Without a loss of generality we make an assertion:

$s = p(\mu_i; \lambda_j \mid i \in \mathcal{S}_p, j \in \mathcal{G}_p, C)$ and

$t = q(\mu_i; \lambda_j \mid i \in \mathcal{S}_q, j \in \mathcal{G}_q, (\exists\, k \in \mathcal{S}_q^{oc} \cup \mathcal{G}_q^{oc})\, p \in \{\mu_k, \lambda_k\}, C)$.

Let $r$ be in A a redex possessing rule preform as a part of a known memory solution RNS for A. Next we construct the following rule preforms:

- micro($r$):

left side: apex(left(micro($r$)))$\mathbb{W}_1$ = apex(left($r$))  and  micro($r$) matches K

right side : First we define notion net induced AlpANBH: Let t be a net. N-AlpANBH is entitled *t-induced*, denoted N-AlpANBH$^I$(t), if N = $\alpha$(rank(t)) is a cover of a net homomorphism image of t, where $\alpha$ is a bijection from the rank alphabet to a set of jungles. Because for each [NUO(t)]-class representative AlpANBH-images are equal for the same AlpANBH, we can now choose right(micro($r$)) to be a net in the preimage domain of an apex(right($r$))-induced AlpANBH. From the same reasons for the case "$r$ is an instance sensitive INRNS-rule", we can choose for each manoeuvre letter x in the domain of right side substitution g of $r$ the x-image of right side substitution of micro($r$) to be a net in the preimage domain of an AlpANBH$^I$(g(x)).

Therefore there is such an AlpANBH, say $\mathbb{W}_{o1}$, that s$\mathbb{W}_1 r$ = K(micro($r$))$\mathbb{W}_{o1}$.

- macro(micro($r$)) (=$\{r_1, r_2\}$) :

For each jungle D we introduce D-induced partition PI(D) :

$$\{\cap D' \setminus \{\cap D'' : D' \subset D'', D'' \in P(D)\} : D' \in P(D)\}$$

P indicating the power set of its argument. We construct such a new AlpANBH, say $\mathbb{W}_3$ : K$\mapsto$B that block($\mathbb{W}_3$) is a partition induced by union block($\mathbb{W}_1$)$\cup$block($\mathbb{W}_2$):

1. first executed rule preform (*ferp*) $r_1$ : t$\mathbb{W}_2 \mapsto$(NUO$^{-1}$(t))$\mathbb{W}_3$, the choice guaranteeing in rewriting processes perseverance of analogous environments in K$\mathbb{W}_2$ and K.

2. secondly executed rule preform (*serp*) $r_2$ :

left side: apex(left($r_2$)) = (NUO$^{-1}$(apex(left(micro($r$)))))$\mathbb{W}_3$  and  $r_2$ matches (NUO$^{-1}$(t))$\mathbb{W}_3$

right side: Following analogously constructing right side of micro($r$) we are free to choose an intervening AlpANBH, say $\mathbb{W}_{o2}$, with the block equal with the block of $\mathbb{W}_{o1}$, and accordingly we choose: apex(right(micro($r$)))$\mathbb{W}_{o2}$ = apex(right($r_2$)).

Therefore $t\mathcal{W}_2\, r_1\, r_2 = K(\text{micro}(r))\mathcal{W}_{o2}$. We have reach:

RESULT 2.1. For each RNS $\mathscr{R}$ there is RNS $\mathscr{P} = \{\text{ferp,serp}\}$ satisfying commutative equation $\theta\mathscr{R} = \mathscr{P}\theta$ for abstraction relation $\theta$ of type T in ITG. We denoted $\mathscr{P} \leadsto^{f}_{\theta} \mathscr{R}$ and call those RNS´s $\theta$-*fixed parallel* RNS´s, $\mathscr{P}$ fixed to $\mathscr{R}$. Generalization of $\leadsto^{f}_{\theta}$ for TD´s is orchestrated with accordance to relation $\leadsto^{c}_{\theta}$. If $\theta$ has no requirements in addition to its type, say T, we denote $\leadsto^{f}_{T}$ for this *fixed parallel TD-relation* or simply $\leadsto^{f}$. Clearly $\leadsto^{f}_{T}$ is a genuine subset of $\leadsto^{c}_{T}$.

For each *renetting system* $\mathscr{R}$ there is $\theta$-*fixed parallel* RNS $\mathscr{P} = \{\text{ferp}_{\mathscr{R},\mathscr{P}}, \text{serp}\}$ ($\mathscr{P}$ fixed to $\mathscr{R}$) satisfying commutative equation $\theta\mathscr{R} = \mathscr{P}\theta$ for abstraction relation $\theta$ of type T in ITG, Tirri SI (2013 May) denoted $\mathscr{P} \leadsto^{f}_{\theta} \mathscr{R}$. Generalization of $\leadsto^{f}_{\theta}$ for TD´s, *fixed parallel TD-relation*, is orchestrated with accordance to relation $\leadsto^{c}_{\theta}$. If $\theta$ has no requirements in addition to its type it is not indicated. Due to the lack of reflexivity $\leadsto^{f}_{\theta}$ is a genuine subset of $\leadsto^{c}_{\theta}$ with cardinality $|\leadsto^{c}_{\theta}| < 2|\leadsto^{f}_{\theta}|$.

Renetting systems in the same element of

$\text{En}(A,\theta) = \{\ B\subseteq A : \cap(\mathscr{R}\leadsto^{f}_{\theta} : \mathscr{R}\in B) \neq \varnothing,\ |B| > 1\ \}$,

where A is a sample of RNS´s and $\theta\in$ ITG, constitute pairwise *entangling equivalence relation* $\overset{E}{\leadsto}_{\theta}$ in the set of all RNS´s, if A is establishing the set of all RNS´s. We recall result: $\overset{E}{\leadsto}_{\theta} = \leadsto^{c}_{\theta}$ saying $\leadsto^{f}_{\theta}$ *generates* $\leadsto^{c}_{\theta}$ Tirri SI (2014 Jan). We agree $\mathscr{F}$ *fundamental TD set* standing for the set of the TD´s and define *entangling mediatory set for* each $B\in\text{En}(\mathscr{F},\theta)$

$\mathcal{E}t(B) = \cap(\mathscr{R}\leadsto^{f}_{\theta} : \mathscr{R}\in B)$.

Clearly $\leadsto^{f}_{\theta}$ is a genuine subset of $\leadsto^{c}_{\theta}$. Consequently generalizations of entangling relation and mediatory set for the set of TD´s are presumed to be established as for parallel relation.

## HERMENEUTIC AREA

**Definition 2.1.** PARTICIPATORY TD. Let $\theta \in$ ITG and $\mathcal{R} \in \mathcal{F}$. We say

$$\mathcal{R}(\mathcal{R}) = \{ \, \mathcal{P} : \mathcal{P} \leadsto \mathsf{f}_\theta \,=\, \mathcal{R} \leadsto \mathsf{f}_\theta \, \},$$

is $\mathcal{R} \leadsto \mathsf{f}_\theta$ -class *generating* set of TD´s,

$$\wp(\mathcal{R}) = \{ \, \mathcal{P} : \text{nest}(\mathcal{P}) = \{ \, \mathscr{R} : r \text{ is of serp-type, } r \in \mathscr{R} \, \}, \, \mathcal{P} \in \mathcal{R}(\mathcal{R}) \, \}$$

is *the center set of* $\mathcal{R} \leadsto \mathsf{f}_\theta$ -class manifesting *participatory operation* $\wp$ in $\mathcal{F}$ and $\wp(\mathcal{F})$ is representing "participatory TD family" of $\leadsto \mathsf{f}_\theta$ .

**Definition 2.2.** FERP GENERATION. For each TD $\mathcal{R}$ we define *Cartesian ferp generation set from TD* $\mathcal{R}$

$$\mathcal{R} \leadsto^{\mathsf{ferp}} = \{ \, ( \Pi(\text{ferp}_{x,y} : y \in A_x) : A_x \in P(\mathcal{R} \leadsto \mathsf{f}_\theta), \, x \in \mathcal{R} \leadsto \mathsf{f}_\theta \, \},$$

where $\Pi(.)$ is the notation for Cartesian product over $(.)$ and manifesting transitive closure *ferp generation relation* $\leadsto^{\mathsf{ferp}}$ in $\mathcal{F}$, generating $\leadsto^{\mathsf{c}}_\theta$ and constituting a closure operation in $\mathcal{F} / \leadsto^{\mathsf{c}}_\theta$ .

**Definition 2.3.** HERMENEUTIC AREA. For each TD $\mathcal{R}$ *Hermeneutic area of class* $\mathcal{R} \leadsto^{\mathsf{c}}_\theta$ is

$$\mathcal{H}(\mathcal{R}) = \{ \text{h} : \text{apex}(\text{left}(\text{h})) \in \text{Par}(\text{PI}(\{\text{apex}(\text{left}(\mathcal{R}))\mathcal{W}_1\hat{\ }, \text{apex}(\text{left}(\mathcal{P}))\mathcal{W}_2\hat{\ }\}), \mathcal{W}_1, \mathcal{W}_2 \text{ are of type } \theta \text{ RNS´s}, \mathcal{P} \in \mathcal{R} \leadsto^{\mathsf{c}}_\theta \, ) \},$$

manifesting *hermeneutic operation* $\mathcal{H}$ in $\mathcal{F}$ and where $\mathcal{W}\hat{\ }$ is a normal form of RNS $\mathcal{W}$ and PI is induced partition for its arguments.

ENTANGLING PROPOSITION. Because $\leadsto^{\mathsf{E}}_\theta \,=\, \leadsto^{\mathsf{c}}_\theta$ , $\theta \in$ ITG, for each $\mathcal{R}$ we obtain

$$\mathcal{H}(\mathcal{F} \leadsto^{\mathsf{c}}_\theta) \,=\, \mathcal{E}t(\text{En}(\mathcal{F}, \theta)),$$

$$\mathcal{H}(\mathcal{F} \leadsto^{\mathsf{c}}_\theta) \,\subseteq\, \mathcal{F} \leadsto^{\mathsf{c}}_\theta \,=\, \wp(\mathcal{F}) \leadsto^{\mathsf{ferp}},$$

$$\mathcal{H}(\mathcal{R}) \text{ is a fixed point of operation } \leadsto^{\mathsf{ferp}\hat{\ }}$$

and $\wp(\mathcal{F})$ is a minimal generating set of hermeneutic area of all $\leftrightsquigarrow_{\theta}^{c}$-classes subject to transitive closure ferp relation.

**Definition 2.4.** Let $\bar{\Theta}_o \subseteq Eq(F_\Sigma(X,\Xi))$, *fundamental relation set*. For each $k \in \mathbb{N}$ and $t \in F_\Sigma(X,\Xi))$ we denote k-level saturation set of t

$$S_k(t) = \{\{s \in \theta F_\Sigma(X,\Xi) : \theta \in \bar{\Theta}_{(k-1)} \} : s \in F_\Sigma(X,\Xi)\} \cap Sat(t),$$

where *multidimensional abstraction relation set*, $\bar{\Theta}_k = \{\bar{\Theta}_{ki} : i = 1,\ldots, i_k, C_k\}$, $i_k \in \mathbb{N}$ , is defined as the direct product of elements in $\bar{\Theta}_{(k-1)}$ such that for each s and t in $F_\Sigma(X,\Xi)$ $(s,t) \in \bar{\Theta}_{ki}$ , i= 1,…, $i_k$ , if and only if

$$\exists (P \in S_k(s), Q \in S_k(t)) \quad (\forall p \in P) (\exists q \in Q) (\exists \theta \in \text{л}(k\text{-}1,i)) \quad (p,q) \in \theta,$$ where л$(k\text{-}1,i) \subseteq \bar{\Theta}_{(k-1)}$,

and revised for q and p respectively; $C_k$ being *boundary condition set*, establishing qualifications for applied algebras. For each net t Cartesian $\prod(|S_n(t)|: n \in \mathbb{N})$ indicates *Gravity of* t, *grav*(t), the complexity rate of relation folding.

We recall definitions and notations of the following:

*Generalized parallel TD-relation* (GPTDR) subject to *multidimensional abstraction relation set*,

*TD-л-abstraction relation* $\theta_{TDл}$ , л$\in$ITG, where for л-derived TD-operations H $\theta_{TDл}$ K,

if and only if $H^{-\mathcal{N}} \theta_{Nл} K^{-\mathcal{N}}$, where $\mathcal{N}$ stands for *renetting algebra*, $\theta_{Nл}$ is the л-abstraction relation in the set of the nets and TD$^{-\mathcal{N}}$ stands for the carrier net of TD in concern, and

*First order abstraction algebra* (partially quotient algebra) $(A\bar{\Theta}_k, \mathcal{F} \leftrightsquigarrow_{\bar{\Theta}_k})$ with A a jungle, $\mathcal{F}$ a fundamental TD set, $\bar{\Theta}_k$ multidimensional abstraction relation in A at k-dimension $(k \in \mathbb{N}_o)$ subject to $\bar{\Theta}_o$ being a singleton comprising $\theta_N (\in$ITG) and $\leftrightsquigarrow_{\bar{\Theta}_k}$ GPTDR in $\mathcal{F}$ subject to $\bar{\Theta}_k$ .

**N-th ORDER ABSTRACTION ALGEBRA.** Because $\theta_{TD}$ is saturated by $\leftrightsquigarrow_{\theta_N}$ , we define for each $k \in \mathbb{N}_o$ quotient relation in fundamental TD set $\mathcal{F}$, *second order abstraction relation* $\bar{\Theta}_{kTD} / \leftrightsquigarrow_{\bar{\Theta}_k}$ manifesting

$$f \leftrightsquigarrow_{\bar{\Theta}_k} \bar{\Theta}_{kTD} / \leftrightsquigarrow_{\bar{\Theta}_k} g \leftrightsquigarrow_{\bar{\Theta}_k}, \quad \text{if } f \bar{\Theta}_{kTD} g.$$

and *second order parallel relation* in $P(\mathcal{F})$, $\infty_{(\bar{\Theta}_{k\text{TD}}\,/\,\infty\,\bar{\Theta}_k)}$ :

$$(\forall\, S,T \subseteq \mathcal{F})\;\; S\;\infty_{(\bar{\Theta}_{k\text{TD}}\,/\,\infty\,\bar{\Theta}_k)}\;T\;, \text{ if}$$

$$(f\infty_{\bar{\Theta}_k})\odot S\;\;\bar{\Theta}_{k\text{TD}}\,/\infty\,\bar{\Theta}_k\;\;(g\infty_{\bar{\Theta}_k})\odot T\;, \text{ whenever }\;(f\infty_{\bar{\Theta}_k}\,,\,g\infty_{\bar{\Theta}_k}) \in \bar{\Theta}_{k\text{TD}}\,/\infty\,\bar{\Theta}_k$$

with *transducer reversing binary operation* $\odot$ in $\mathcal{F}$: $s\odot t\;=\;(s^{-\mathcal{N}}\,t)^{\mathcal{N}}$.

Setting requisite $\theta_N$ is distinct demonstrates a closure system: *second order abstraction algebra*

$$(\mathcal{F}(\bar{\Theta}_{k\text{TD}}\,/\infty\,\bar{\Theta}_k)\,,\,\odot P(\mathcal{F})\infty_{(\bar{\Theta}_{k\text{TD}}\,/\,\infty\,\bar{\Theta}_k)}).$$

By multiple order abstraction relation enumeration we extend algebra level to n-th order:

For each $n\in\mathbb{N}_0$ and $k_n\in\mathbb{N}_0$ we define relation $\bar{\Theta}_{n,k_n\text{TD}}$ in $P^{n-1}(\mathcal{F})$ :

$$\bar{\Theta}_{0,k_0\text{TD}}\;=\;\bar{\Theta}_{k_0\text{TD}}\;\text{ and}$$

$$(\forall\, H,K\in P^{n-1}(\mathcal{F}))\;\; H\;\bar{\Theta}_{n,k_n\text{TD}}\;K, \text{ if }\;(\cup^n(H))^{-\mathcal{N}}\;\;\bar{\Theta}_{n-1,k_{n-1}\text{TD}}\;\;(\cup^n(K))^{-\mathcal{N}},$$

where for each $n\in\mathbb{N}_0$ inductively stated $\cup^n(B)=\cup(\cup^{n-1}(B))$.

Furthermore we assume $\bar{\Theta}_{n,k_n\text{TD}}$ is distinctive and agree with the notations:

$$\langle\theta\,/\infty\rangle^{\langle\,0,k_0\,\rangle}\;=\;\bar{\Theta}_{0,k_0\text{TD}}\;,$$

$$\langle\theta\,/\infty\rangle^{\langle\,n,k_n\,\rangle}\;=\;\bar{\Theta}_{n,k_n\text{TD}}\,/\infty_{\langle\theta\,/\infty\rangle^{\langle\,n-1,k_{n-1}\,\rangle}}\;.$$

For power sets $T_n\in P^n(\mathcal{F})$, $n\in\mathbb{N}_0$ and $S_n\;=\;S_{n-1}\cup T_n$, $n\in\mathbb{N}$, $S_o=T_o$

$$\mathfrak{R}^{\langle 0\rangle}(S_o,k_o)\;=\;T_o\bar{\Theta}_{0,k_0\text{TD}}$$

$$\mathfrak{R}^{\langle n\rangle}(S_n,k_n)\;=\;\mathfrak{R}^{\langle n-1\rangle}(S_{n-1},k_{n-1})\langle\theta\,/\infty\rangle^{\langle\,n,k_n\,\rangle}\;\odot\;T_n\infty_{\langle\theta\,/\infty\rangle^{\langle\,n,k_n\,\rangle}}\in\{B\subseteq k\in q\langle\theta\,/\infty\rangle^{\langle\,n,k_n\,\rangle}: q\in P^{n-1}(\mathcal{F})\}.$$

We define *n-th order $k_n$-dimensional partially quotient abstraction algebra*

$$(\;\mathfrak{R}^{\langle n-1\rangle}(\;\cup(P^i(\mathcal{F}):i=0,1,...,n-1),k_{n-1})\langle\theta\,/\infty\rangle^{\langle\,n,k_n\,\rangle}\,,\,\odot P^n(\mathcal{F})\infty_{\langle\theta\,/\infty\rangle^{\langle\,n,k_n\,\rangle}}\;).$$

Because mutual gravity value for realities of parallel type guarantees equal partition element numbers yielding common abstraction classes for ITG-type fundamental relation set Tirri SI (2014 Jan) , due to entangling propotion we reach: the elements in *multi-algebra set*

$$\{(\Re^{\langle i-1\rangle}(\cup(\mathcal{H}(P^k(\mathcal{F})):k=0,1,\dots,i-1)\,,\,\mathfrak{z}(i-1)\rangle\langle\theta(\bar{\varrho})\,/\infty\rangle^{\langle i,\mathfrak{z}(i)\rangle}\,,\,\odot\mathcal{H}(P^i(\mathcal{F}))\infty_{\langle\theta(\bar{\varrho})\,/\infty\rangle}^{\langle i,\mathfrak{z}(i)\rangle}):i\in\mathbb{N},\,\mathfrak{z}\in\Psi\}$$

equalizes with the elements in *Participatory algebra set*

$$\{(\Re^{\langle i-1\rangle}(\cup(\wp(P^k(\mathcal{F})):k=0,1,\dots,i-1)\,,\,\mathfrak{z}(i-1)\rangle\langle\theta(\bar{\varrho})\,/\infty^{\mathrm{ferp}\wedge}\rangle^{\langle i,\mathfrak{z}(i)\rangle}\,,\,\odot\,\wp(P^i(\mathcal{F}))\infty^{\mathrm{ferp}\wedge}_{\langle\theta(\bar{\varrho})\,/\infty^{\mathrm{ferp}\wedge}\rangle^{\langle i,\mathfrak{z}(i)\rangle}}):i\in\mathbb{N},\,\mathfrak{z}\in\Psi\}$$

generating $(\mathcal{F},\theta(\bar{\varrho}))$-*information space modulo abstraction relation* $\theta(\bar{\varrho})$, where $\mathfrak{z}$ ( i)-level $\theta$ subject to fundamental relation set $\bar{\varrho}$ representing "*horizontal*" $\mathfrak{z}(i)$-dimensional structure of elements is liable to i-th order "*vertical*" process iteration and $\Psi$ is the set of mappings in $\mathbb{N}$ , hermeneutic $\mathcal{H}$ and participatory $\wp$ being understood as presenting class operations and $\infty^{\mathrm{ferp}}$ quotient relation. We have reached

**Result 2.2.** For each $i\in\mathbb{N}$ and $\mathfrak{z}\in\Psi$

$$\langle\theta(\bar{\varrho})\,/\infty^{\mathrm{ferp}\wedge}\rangle^{\langle i,\mathfrak{z}(i)\rangle}\quad\text{and}\quad\odot\wp(P^i(\mathcal{F}))\infty^{\mathrm{ferp}\wedge}_{\langle\theta(\bar{\varrho})\,/\infty^{\mathrm{ferp}\wedge}\rangle^{\langle i,\mathfrak{z}(i)\rangle}}$$

commute with each other, an intrinsic character of Participatory algebra.

## 3. §     Multi-Dimensional Confluence in Participatory Algebra

Let $\mathcal{G}=\{\mathcal{G}(i,\mathfrak{z}):i\in\mathbb{N},\,\mathfrak{z}\in\Psi\}$, *information space*, stands for the set of the elements in Participatory algebra, abstraction relation $\theta(\bar{\varrho})$ related to quotient relation $\infty^{\mathrm{ferp}}$. Let $\mathcal{D}$ be a family of *dimensions* i.e. a set of quotient jungles ($\subseteq\mathcal{G}$) distinct from each other. Let $h_n$ be of n-th grade AlpANBH, $n\in\mathbb{N}$ and $\mathbb{H}=\{\,h_n:n\in\mathbb{N}\,\}$. Because elements in $\mathbb{H}$ can be seen as special cases of TD´s SI Tirri (2013 Aug), we can apply operation $\odot$ on them establishing topologic metrics in information space via net block homomorphism. For each $\mathcal{D}_o\subseteq\mathcal{D}$ we define *initial* family ($\mathcal{D}_o$-*dimensional* $\mathcal{G}$-*elements*)

$$T^{\mathcal{G}}_{\mathbb{H},\mathcal{D}o}=\{\,s:(\cup\mathcal{D}_o)^{\neg\mathcal{N}}\cap(s\odot h)^{\neg\mathcal{N}}\in\mathrm{Par}((s\odot h)^{\neg\mathcal{N}}),\,s\in\mathcal{G},\,h\in\mathbb{H}\,\},$$

where Par is indicating to the set of the partitions of its argument and $\cap$ stands for overlapping,

and we define furthermore *dimensional closure preserving set* of TD´s

$$\mathcal{R}_{\mathcal{D}_o} = \{\, r : (\mathrm{enc}((\cup E)^{\neg \mathcal{N}} h^{-1}) r \circ h)^{\neg \mathcal{N}} \subseteq \mathrm{enc}((\cup E)^{\neg \mathcal{N}})\,,\ r \in \mathcal{F}_o,\ h \in \mathbb{H}\,,\ E \subseteq \mathcal{D}_o\,\},$$

where $\mathcal{F}_o$ is a set of the rules in the RNS´s of type UPRNS excluding INRNS of TD´s in fundamental TD set $\mathcal{F}$ in $\mathcal{G}$ (Tirri SI 2013 May, Aug) matching only the nets of the sets in family $D_o^{\neg \mathcal{N}} h^{-1}$. We denote *dimensional rewriting operation set* $\mathbb{R} = \{\mathcal{R}_{\mathcal{D}_o} : \mathcal{D}_o \subseteq \mathcal{D}\}$. Let then $v \in \mathcal{G}$ be a *ground element* and $D_a, D_b \subseteq \mathcal{D}$. Let $r_a$ and $r_b$ be two operations in $\mathcal{G}$ of the same vertical and horisontal level. We define *connector set of* $(r_a, r_b)$, $\mathrm{con}(r_a, r_b) \subseteq (\mathcal{R}_{D_a}, \mathcal{R}_{D_b})$, and denote generally $(\mathrm{cn}(r_a), \mathrm{cn}(r_b)) = \mathrm{con}(r_a, r_b)$.

We say that pair $(r_a, r_b)$ is $(D_a \cup D_b)$-*dimensional confluence* from v, if for each $(r_{ad}, r_{bd}) \in (\mathrm{cn}(r_a), \mathrm{cn}(r_b))$

$$\mathrm{enc}((v \circ r_a r_{bd})^{\neg \mathcal{N}}) \subseteq \mathrm{enc}((v \circ r_b r_{ad})^{\neg \mathcal{N}})\ \text{ or }\ \mathrm{enc}((v \circ r_a r_{bd})^{\neg \mathcal{N}}) \supseteq \mathrm{enc}((v \circ r_b r_{ad})^{\neg \mathcal{N}})\,,$$

where proper inclusions indicate *partial confluence*, otherwise confluence is *comprehensive*.
We obtain the result:

Theorem 3.1. Pair $(r_a, r_b)$ is $(D_a \cup D_b)$-dimensional confluence from v, if $\mathrm{cn}(r_a)$ and $\mathrm{cn}(r_b)$ satisfy the following conditions:

- Demands for each $r_{bd} \in \mathrm{cn}(r_b)$ :

$$\mathrm{enc}((v \circ r_b)^{\neg \mathcal{N}}) \cap T^{\mathcal{G}}_{\mathbb{H}, D_b}{}^{\neg \mathcal{N}} \subseteq \mathrm{enc}((v \circ r_a r_{bd})^{\neg \mathcal{N}}) \cap T^{\mathcal{G}}_{\mathbb{H}, D_b}{}^{\neg \mathcal{N}}$$

and the nets of $\mathrm{enc}((v \circ r_b)^{\neg \mathcal{N}}) \cap T^{\mathcal{G}}_{\mathbb{H}, D_b}{}^{\neg \mathcal{N}}$ have no outward connections to any net of

$\mathrm{enc}((v \circ r_a r_{bd})^{\neg \mathcal{N}}) \cap T^{\mathcal{G}}_{\mathbb{H}, D_b}{}^{\neg \mathcal{N}}$, but are linked to nets of $\mathrm{enc}((v \circ r_a r_{bd})^{\neg \mathcal{N}}) \cap T^{\mathcal{G}}_{\mathbb{H}, D_a}{}^{\neg \mathcal{N}}$.

- Demands for each $r_{ad} \in \mathrm{cn}(r_a)$ :

1. $\mathrm{enc}((v \circ r_b)^{\neg \mathcal{N}}) \cap T^{\mathcal{G}}_{\mathbb{H}, D_a}{}^{\neg \mathcal{N}} \subseteq \mathrm{enc}((\mathrm{enc}((v \circ r_a r_{bd})^{\neg \mathcal{N}}) \cap T^{\mathcal{G}}_{\mathbb{H}, D_b}{}^{\neg \mathcal{N}}) r_{ad})$  or

2. (i) $\mathrm{enc}(\mathrm{enc}((v \circ r_a r_{bd})^{\neg \mathcal{N}}) \cap T^{\mathcal{G}}_{\mathbb{H}, D_b}{}^{\neg \mathcal{N}}) \supseteq (\mathrm{enc}((v \circ r_b)^{\neg \mathcal{N}}) \cap T^{\mathcal{G}}_{\mathbb{H}, D_b}{}^{\neg \mathcal{N}}) r_{ad}$ or

   (ii) $\mathrm{enc}((v \circ r_a r_{bd})^{\neg \mathcal{N}}) \cap T^{\mathcal{G}}_{\mathbb{H}, D_b}{}^{\neg \mathcal{N}} \subseteq \mathrm{enc}((\mathrm{enc}((v \circ r_b)^{\neg \mathcal{N}}) \cap T^{\mathcal{G}}_{\mathbb{H}, D_b}{}^{\neg \mathcal{N}}) r_{ad})$.

If the inclusions in demands for $r_{ad}$ and $r_{bd}$ are proper, partial confluence is guaranteed; otherwise confluence is comprehensive. Furthermore we could construct for the families of any partition of set $(D_a \cup D_b)$ dimensional rewriting catenations satisfying the demands requisite for $cn(r_a)$ and $cn(r_b)$ thus guaranteeing $(D_a \cup D_b)$-*dimensional confluence* for pair $(r_a, r_b)$.

However it is not possible to avoid usage of all dimensions in $D_a \cup D_b$ expressed in applying pair $(r_a, r_b)$ to v in respect to achieve confluence via dimensional rewriting $\mathbb{R}$, if there exist the following conditions in the results of pair $(r_a, r_b)$ :

First let $r \in \mathcal{R}_{D_a} \cup \mathcal{R}_{D_b}$ : We separate two possible cases (i) and (ii):

(i) $enc((v \odot r_a)^{-\mathcal{N}} r) \supseteq (v \odot r_b)^{-\mathcal{N}}$

is not possible, if

$enc((v \odot r_b)^{-\mathcal{N}}) \cap T^{\mathcal{G}}_{\mathbb{H}, D_a}{}^{-\mathcal{N}} \nsubseteq enc((enc((v \odot r_a)^{-\mathcal{N}}) \cap T^{\mathcal{G}}_{\mathbb{H}, D_a}{}^{-\mathcal{N}})$

and $enc((v \odot r_b)^{-\mathcal{N}}) \cap T^{\mathcal{G}}_{\mathbb{H}, D_b}{}^{-\mathcal{N}} \nsubseteq enc((enc((v \odot r_a)^{-\mathcal{N}}) \cap T^{\mathcal{G}}_{\mathbb{H}, D_b}{}^{-\mathcal{N}})$

(ii) $enc((v \odot r_b)^{-\mathcal{N}} r) \supseteq (v \odot r_a)^{-\mathcal{N}}$

is not possible, if

$enc((v \odot r_a)^{-\mathcal{N}}) \cap T^{\mathcal{G}}_{\mathbb{H}, D_b}{}^{-\mathcal{N}} \nsubseteq enc((enc((v \odot r_b)^{-\mathcal{N}}) \cap T^{\mathcal{G}}_{\mathbb{H}, D_b}{}^{-\mathcal{N}})$

and $enc((v \odot r_a)^{-\mathcal{N}}) \cap T^{\mathcal{G}}_{\mathbb{H}, D_a}{}^{-\mathcal{N}} \nsubseteq enc((enc((v \odot r_b)^{-\mathcal{N}}) \cap T^{\mathcal{G}}_{\mathbb{H}, D_a}{}^{-\mathcal{N}})$.

## Summarizing confluence prerequisites.

In general for each distinct dimension set matched by an unconstrained operation there has to be a constrained dimensional operation matching the set, if that operation is of type UPRNS excluding INRNS.

**Definitions 3.1.** Let $i \in \mathbb{N}$, $\mathfrak{z} \in \Psi$ and $S \subseteq \mathcal{G}(i, \mathfrak{z})$, $|S| \in \mathbb{N}_0$. We say that pair $(v, S)$ is $S(i, \mathfrak{z})$-*ranked dimensional confluent*, if there is such a dimensional rewriting operation set of $\mathcal{G}(i, \mathfrak{z})$, say L, that family $v \odot SL$ is a singleton and we entitle SL as $(v, S)$-*harmonizer*. It is to be notified that the limit

demands in RNS´s of L are a matter of the uttermost importance albeit easily adjustable case by case. For each $r \in \mathcal{G}(i,\check{\sigma})$, the set of the catenations of the rule preforms of the RNS´s in $r$ we denote CatRp($r$) and as accustomed we agree with CatRp(S) = $\cup$(CatRp($r$) : $r \in$S). For each K$\subseteq \mathcal{G}(i,\check{\sigma})$, we denote con(K$^2$) = $\cup$(con($r_1$, $r_2$) : ($r_1$, $r_2$)$\in$K$^2$). Let $\mathscr{E}$ be a mapping : K $\mapsto$ the set of the elements of the pairs in con(K$^2$) and $\mathfrak{C}$(S) = $\cup$($\mathscr{E}^q$( CatRp(S)) : q$\in$IN$_0$ ) , where $\mathscr{E}^0$ is the identity mapping. From theorem 3.1 applied to comprehensive connectors we obtain the following result:

Theorem 3.2. There is a subset of $(\mathfrak{C}(\mathrm{S}))^2$ of comprehensive connectors, say T, guaranteeing for each i$\in$IN and $\check{\sigma}\in\Psi$ S(i,$\check{\sigma}$)-ranked dimensional confluence. Furthermore the minimum cardinality of T is (|S|-1)·|S|/2, because for each A$\subseteq$S to be rewritten to the set with the reduced cardinality |A|-1, we need at least |A|-1 connectors.

CONCLUSIONS

Confluence in abstract parallel category systems is established for net class-rewriting in iterative closed multilevel quotient graph structures with uncountable node arities by multi-dimensional transducer operations in topological metrics defined by alphabetically abstracting net block homomorphism. We obtain minimum prerequisites for the comprehensive connector pairs in a multitude dimensional rewriting closure generating confluence in participatory algebra for different horizontal and vertical level projections modulo abstraction relation constituting formal semantics for confluence in information space. Due to invariabilities between parallel pairs in class rewriting in Participatory algebra liable to each fundamental relation set $\bar{\varrho}$, horizontal structure mapping $\check{\sigma}$ and vertical process iteration cardinality and due to commutation result 2.2 reached confluences from ground level are established via ferp generation with the adjacent harmonizer´s parallel counterparts in $\wp$ (P$^i$($\mathscr{F}$)) $\infty^{\mathrm{ferp}^\wedge}_{\langle\theta(\bar{\varrho}) / \infty^{\mathrm{ferp}^\wedge}\rangle^{\langle i, \check{\sigma}(i)\rangle}}$ , consequently constituting the closure of all quotient confluences related to prerequisites for parameters in Participatory algebra as a subalgebra, *confluence algebra*, and with formal automata syntax in its entirety achieved via automated problem solving paradigm.

## ACKNOWLEDGEMENTS

I own the unparalleled gratitude to my family, my wife and five children for the cordial environment so very essential on creative working.